\documentclass[11pt]{iopart}

\usepackage{amssymb}
\usepackage{graphicx}

\newcommand{\bee}{\begin{equation}}
\newcommand{\eee}{\end{equation}}
\newcommand{\bea}{\begin{eqnarray}}
\newcommand{\eea}{\end{eqnarray}}
\newcommand{\be}{\begin{equation}}
\newcommand{\ee}{\end{equation}}

\newcommand{\llsim}{\lesssim} 
\newcommand{\har}[1]{^{(#1)}}

\newcommand{\al}{\alpha}
\renewcommand{\b}{\beta}
\newcommand{\de}{\delta} 
 
\newcommand{\ep}{\epsilon} 
\newcommand{\ga}{\gamma}

\newcommand{\om}{\omega}

\newcommand{\veps}{\varepsilon} 

\begin{document}
\title{Constraints on the Electrical Charge Asymmetry of the Universe}
\author{C.~Caprini and P.~G.~Ferreira}
\ead{chiara.caprini@physics.unige.ch}
\address{D\'epartement de Physique Th\'eorique, Universit\'e de
  Gen\`eve, 24 quai Ernest Ansermet, CH--1211 Gen\`eve 4, Switzerland}
\ead{p.ferreira1@physics.ox.ac.uk}
\address{Astrophysics, University of Oxford,
Denys Wilkinson Building, Keble Road, Oxford OX1 3RH, UK}

\begin{abstract}
We use the isotropy of 
the Cosmic Microwave Background to place stringent constraints 
on a possible electrical charge asymmetry of the universe. 
We find the excess charge per baryon to be $q_{e-p}<10^{-26}e$
in the case of a uniform distribution of charge, where $e$ is the charge of
the electron. If the charge asymmetry is 
inhomogeneous, the constraints will depend on the spectral index, $n$, of
the induced magnetic field and range from $q_{e-p}<5\times 10^{-20}e$ ($n=-2$)
to $q_{e-p}<2\times 10^{-26}e$ ($n\geq 2$). 
If one could further assume that the charge asymmetries of individual particle species
are not anti-correlated so as to cancel, this would imply, for photons, 
$q_\gamma< 10^{-35}e$; for neutrinos, $q_\nu<4\times10^{-35}e$; and for
heavy (light) dark matter particles $q_{\rm dm}<4\times10^{-24}e$ 
($q_{\rm dm}<4\times10^{-30}e$).

\end{abstract}

\maketitle

\section{Introduction}

With the substantial improvement of experimental and observational
techniques in particle physics an astrophysics, it has become possible to
test some of the assumptions that go into constructing
models of fundamental interactions and the universe. 
The aim of this work is to place a cosmological constraint on the
presence of an electrical charge asymmetry in the universe.
The first cosmological analysis of a charged universe 
undertaken was in the context of the Steady State model of the
universe by Bondi \& Lyttleton in 1959 \cite{BL}, where an attempt
was made to explain the recession of the galaxies through
electromagnetic repulsion. 
To implement charge non-conservation,
Maxwell's equations were modified to include a direct coupling to
the vector potential, violating gauge invariance. Swann
\cite{Swann1961} confuted this analysis; Barnes \cite{AB} correctly reanalysed
a similar model,
using the Proca theory of electromagnetism. 
The spontaneous breaking of gauge symmetry, and the subsequent development
of a charge imbalance, 
was considered by Ignatiev \emph{et al.} \cite{IKS},
and Dolgov \& Silk \cite{DS}
used it as a mechanism of creation of
primordial magnetic fields. An implementation of a
homogeneous and isotropic charge density was proposed in the
context of massive electrodynamics \cite{BWK}, and the possibility
of charge non-conservation has been analysed in brane world models
\cite{DRT}, and in varying speed of light theories \cite{Landauetal2001}.

In principle, a potential charge asymmetry could be carried by a number
of different, stable components in the universe. These can include: photons, neutrinos,
dark matter, or a difference in the electron and proton charges. 
Experimental constraints, based on the lack of dispersion from 
pulsar signals, limit the photon charge to be $q_\gamma < 5\times10^{-30}e$
\cite{Raffelt94} (a less restrictive laboratory constraint, $q_\gamma <
8\times10^{-17}e$, has also been established \cite{Semertzidis2003}).
The analysis of the luminosity evolution of red giants in globular clusters leads 
to a constraint on the charge of neutrinos (or the charge of any sub-keV particle) of 
$q_\nu < 2\times10^{-14} e$ \cite{Raffelt99}. Direct, laboratory constraints based 
on gas-efflux measurements
\cite{Piccard}, electro-acoustic techniques \cite{Dylla_King}, and Millikan-type 
experiments using steel balls \cite{Marinelli_Morpurgo}, all limit the 
electron-proton charge difference to be $q_{e-p} < 10^{-21} e$. 
For particles generally 
with masses less than $\sim1$ MeV, Big Bang Nucleosynthesis rules out
charges greater than $10^{-10}e$ \cite{Davidson}. Concerning dark
matter, constraints have been established on the fraction of dark
matter due to charged heavy particles of the order of $10^{-5}$ \cite{Verkerk}.   

In this work we will obtain general constraints on a 
possible charge asymmetry, characterised by a uniform or a
stochastic distribution, placing the potential
contribution of individual components in context. 
The premise will be that charge was generated at some very early 
time, but has been conserved in the period in which we are establishing
constraints.

We use Heaviside-Lorentz electromagnetic units with $e=\sqrt{4\pi
\alpha}$, $c=1 =8\pi G$; Greek
indices run from 0 to 3 and Latin indices from 1 to 3. The scale
factor is normalised to $a(\tau_0)=1$ today, where $\tau$ denotes
conformal time.

\section{Model of a charged universe}

In this analysis we use the 1+3 covariant formalism developed in \cite{EL}: 
we consider a general class of homogeneous space-times for which it is possible
to define a preferred velocity field, $u^{\al}$, determining the
fundamental fluid flow lines and satisfying $u^{\al}u_{\al}=-1$.
The existence of this vector field generates a unique splitting of
spacetime, given by the instantaneous three-dimensional rest space
of an observer moving with 4-velocity $u^\al$ and the
one-dimensional space $u^\al$ itself. The metric tensor of this
spacetime, $g_{\al\beta}$ can be written like
$g_{\al\beta}=h_{\al\beta}-u_{\al}u_{\beta}$, where $h_{\al\beta}$ is a
projection tensor and represents effectively the spatial metric
for the observer moving with $u^\al$. Any physical tensor field on
this spacetime can be separated with respect to $h_{\al\beta}$ and
$u^\al$ into space and time parts. The covariant derivative of a tensor field
$S^{\al\beta}$ splits
into a comoving time derivative $\dot{S}^{\al\b}=u^\ga
{S^{\al\b}}_{;\ga}$ and a covariant spatial
derivative $D_{\ga}S^{\al\b}= {h_\ga}^\de {h^\al}_\mu {h^\b}_\nu
{S^{\mu\nu}}_{;\de}$. In particular, the splitting of the covariant
derivative of the 4-velocity takes the form:
\begin{eqnarray}
u_{\al;\beta}=\sigma_{\al\beta}+\omega_{\al\beta}+\frac{1}{3}\Theta h_{\al\beta}
-{\dot u}_{\al}u_{\beta}\,,
\end{eqnarray}
where $\sigma_{\al\beta}$ is the shear tensor, $\omega_{\al\beta}$ is the
vorticity tensor, $\Theta=3 \dot a /a$ is the volume expansion, and ${\dot
  u}_{\al}=u_{\al;\b}u^{\beta}$ is the
acceleration vector. The electromagnetic field tensor 
$F_{\al\b}=-F_{\b\al}$ is split into electric
and magnetic fields as measured by an observer moving with $u^\al$:
\bee
F_{\al\b}=2u_{[\al} E_{\b]}+\ep_{\al\b\ga}B^\ga\,,
\label{emfieldtensor}
\eee
where the electric and magnetic 3-vectors are given by 
$E_\al=F_{\al\b}u^\b$, $B_\al=\frac{1}{2}\,\ep_{\al\b\ga}F^{\b\ga}$,
and $\ep_{\al\b\ga}=\eta_{\al\b\ga\de}u^\de$ is the volume element for
the orthogonal rest space $h_{\al\beta}$. We define the
current density 4-vector $j^\al$, where $q=-j^\al u_\al$ is the charge
density, and $J^\al={h^\al}_\b j^\b$ is the spatial current measured by
$u^\al$. Then, Maxwell's equations 
${F^{\al\b}}_{;\b}=j^\al$, $F_{\al\b;\ga}+F_{\ga\al;\b}+F_{\b\ga;\al}=0$
become \cite{EL,Maartens_Bassett1998}
\bea 
D_\al E^\al+2\omega_\al B^\al&=&q \label{maxwella}\\
D_\al B^\al-2\omega_\al E^\al&=&0 \label{maxwellb}\\
h_{\al\b} \dot{E}^\b-\ep_{\al\b\ga}D^\b B^\ga&=&
-J_\al+\ep_{\al\b\ga}\dot{u}^\b B^\ga+\left(\omega_{\al\b}+\sigma_{\al\b}-\frac{2}{3}\Theta
h_{\al\b}\right)E^\b \label{maxwellc}\\
h_{\al\b} \dot{B}^\b+\ep_{\al\b\ga}D^\b E^\ga&=&
-\ep_{\al\b\ga}\dot{u}^\b E^\ga+\left(\omega_{\al\b}+\sigma_{\al\b}-\frac{2}{3}\Theta
h_{\al\b}\right)B^\b \label{maxwelld}\,,
\eea
where $\om_{\al}=\frac{1}{2}\,\ep_{\al\b\ga} \om^{\b\ga}$ is the vorticity
vector. It appears that the motion of the
observer affects the form of Maxwell's equations: beside the usual
spatial divergences, curls and time derivatives
of the fields, one has the appearance of extra terms representing the
motion of the family of fundamental observers moving with 4-velocity $u^\al$.
Moreover, the current conservation equation
${j^\al}_{;\al}=0$ takes the form
\be
\dot{q}+\Theta q+D_\al J^\al+\dot{u}_\al J^\al=0\,, 
\label{ohmlaw}
\ee
and the covariant form of Ohm's law is
\be
j_\al+u_\al u_\b j^\b =\sigma F_{\al\b}u^\b\,,
\ee
where $\sigma$ denotes the conductivity. Projecting onto the
instantaneous rest space of the fundamental observer, Ohm's law 
becomes $J_\al=\sigma E_\al$.

We consider a model of the universe filled with a charged perfect
fluid, which could be matter or radiation, characterised by the
energy-momentum tensor $T^{\al\b}_{F}=\rho\, u^\al u^\b+p
h^{\al\b}$; the charge will give rise to an electromagnetic field,
with $T^{\al\b}_{EM}={F^\al}_\ga
F^{\b\ga}-\frac{1}{4}g^{\al\b}F_{\ga\de}F^{\ga\de}$. The
energy-momentum conservation equation accounts for the interaction
between the fluid and the field, so that \cite{Jackson1975}
\be
{T_{F}^{\al\b}}_{;\b}=F^{\al\b}j_\b\,.
\ee
Using the above definitions, the
coupling term can be rewritten as
\be
F^{\al\b}j_\b=qE^\al+F^{\al\b}J_\b=qE^\al+u^\al E^\b J_\b+\ep^{\al\b\ga}J_\b B_\ga\,.
\ee
We further assume the usual equation of state for a perfect fluid $p=w\rho$; 
then, the energy conservation equation for the fluid, $u_\al {T^{\al\b}_{F}}_{;\b}=u_\al
F^{\al\b}j_\b$, takes the form 
\be
\dot\rho+(1+w)\Theta\rho=-E^\b J_\b\,,
\ee
and the momentum conservation equation $h_{\al\b} {T^{\b\ga}_{F}}_{;\ga}=h_{\al\b}
F^{\b\ga}j_\ga$ takes the form
\be
(1+w)\rho\,\dot{u}_\al+w D_\al \rho =qE_\al+\ep_{\al\b\ga}J^\b B^\ga\,.
\ee
If the charged fluid is composed by non-relativistic matter, one
defines 
$q=\veps \,e\,n_{\rm mat}=\frac{\veps\,e}{m}\rho_{\rm mat}$, where $\veps\,e$
and $n_{\rm mat}$ represent respectively the charge and the number density of the fluid
particles, and $m$ is their mass; 
if instead the radiation component is charged, one
has $q=\veps \,e\,n_{\rm rad}\propto \rho_{\rm rad}^{3/4}$, with similar definitions. 
It appears from this picture that the introduction of a charged 
fluid breaks the isotropy of the universe, through the creation of
currents and electromagnetic fields. Therefore, it is in principle possible to use
the CMB as a measure of the isotropy of the universe to constrain the presence
of an overall charge asymmetry. 

We must further consider the fact that the universe has a very
high conductivity, during all the phases of its
evolution. The conductivity varies in the different epochs of the evolution of
the universe,
depending on the number of charge carriers and on the dominant
scattering process. Before $e^+e^-$ annihilation, the conductivity has
been evaluated in \cite{AE}, and is found to decay with temperature
like $\sigma\sim T/(\al^2\log (1/\al))$, where $\al$ is the fine structure
constant. During the radiation dominated era, considering Thomson
scattering and proceeding as in 
\cite{Turnerandwidrow1988}, one finds 
$\sigma=2\pi\frac{n_e}{n_\ga}\frac{m_e}{e^2}\simeq 3\times10^{13}\sec^{-1}$.
 During the matter dominated epoch, the
dominant scattering process is again 
Thomson scattering of CMB photons with the residual free
electrons, and one has $\sigma\sim
3\times10^{10}\sec^{-1}$ \cite{GR}.  
We perform our analysis in the ideal magnetohydrodynamics limit, for which the
conductivity goes to infinity while the current remains finite \cite{Jackson1975}.
In the reference frame of comoving observers, corresponding to
the reference frame of the fluid in this picture, Ohm's law takes
the form $J_\al =\sigma E_\al$. 
Consequently, applying the magnetohydrodynamic limit, 
one finds that the electric field (but not the magnetic field) must go
to zero \cite{TB}. This ensures that the spatial
current remains finite, avoiding the possibility of an 
instantaneous response of the fluid to the electromagnetic fields.  
Note that this is valid because our analysis is performed in the
reference frame of observers comoving with the fluid, characterised by
the 4-velocity $u^\al$; in a reference
frame with 4-velocity $\tilde{u}^\al=u^\al-v^\al$, the
magnetohydrodynamic limit implies the presence of an electric field
$\tilde{E}_\al=-\ep_{\al\b\ga}v^\b \tilde{B}^\ga$. 

Another way of rephrasing the magnetohydrodynamic 
limit is found by considering the time evolution of the electric 
and magnetic fields given by 
Maxwell's equations (\ref{maxwellc}) and (\ref{maxwelld}). 
The presence of a primordial, uniform magnetic field 
in a Friedmann-Robertson-Walker (FRW) universe is accounted for by assuming that the energy density
of the field is a first order quantity in perturbation theory, and
that the field itself is a small background component. We introduce the
presence also of an electric field with the same characteristics. Then,
the evolution of these fields at the lowest order is given by
\bea
\dot{B}_\al&=&-\frac{2}{3}\Theta B_\al \,,\\
\dot{E}_\al&=&-J_\al-\frac{2}{3}\Theta E_\al \,.
\eea
The first of these equations suggests that the magnetic field
varies on an approximative time-scale given by the Hubble time $\Theta^{-1}$. 
By substituting with Ohm's law $J_\al=\sigma E_\al$ in the second equation, one finds that
the time-scale of variation of the electric field is instead of the
order $(\sigma+(2/3)\Theta)^{-1}$. One further has that, for all the
epochs of the evolution of the universe, 
$\sigma/\Theta\gg 1$ (for example, $\sigma/\Theta\sim 10^{28}$
today). Therefore, while the magnetic field
varies on a time-scale comparable to the Hubble time, the electric
field varies on a much smaller time-scale, given by $\sigma^{-1}$, and gets
dissipated much quicker. The infinite conductivity limit provides an
explanation of the reason why large scale magnetic fields, 
and not electric fields, are observed in the universe. 

In the case of a charged cosmic fluid which we are
analysing, and in the reference frame of comoving observers,
adopting the infinite conductivity limit reduces the form of the
relevant equations to 
\bea
&&\dot\rho+(1+w)\Theta\rho=0\label{enconservinfconduct}\,,\\
&&(1+w)\rho\,\dot{u}_\al+w D_\al \rho =\ep_{\al\b\ga}J^\b B^\ga \label{monconservinfconduct}\,,
\eea
for the energy and momentum conservation, and 
\bea 
2\,\omega_\al B^\al&=&q \label{maxwell1infconduct}\\
D_\al B^\al&=&0 \label{maxwell2infconduct}\\
\ep_{\al\b\ga}D^\b B^\ga&=&J_\al-\ep_{\al\b\ga}\dot{u}^\b B^\ga \label{maxwell3infconduct}\\
h_{\al\b} \dot{B}^\b&=&\left(\omega_{\al\b}+\sigma_{\al\b}-\frac{2}{3}\Theta
h_{\al\b}\right)B^\b \label{maxwell4infconduct}\,,
\eea
for Maxwell's equations. We have defined the charge density as 
$q=\veps \,e\,n$, both
in the case of charged non-relativistic matter and radiation;
therefore, using $\rho_{\rm mat}\propto n_{\rm mat}$ and 
$\rho_{\rm rad}\propto n_{\rm rad}^{4/3}$, 
Eq.~(\ref{enconservinfconduct}) implies the
charge evolution $\dot q+\Theta q=0$.
Equation (\ref{maxwell1infconduct}) shows that in a universe with infinite
conductivity, the presence of a charge density implies the
presence of a magnetic field, and induces vorticity in the metric.  
Therefore, the analysis in a non-perturbative framework should be carried on
in the context of tilted Bianchi universes, \emph{i.e.} homogeneous and anisotropic
universes in which the surfaces of homogeneity are not orthogonal to
the matter flow. 
However, the goal of this analysis is to constrain the presence of charge in
the universe by using the isotropy of the CMB. 
Consequently, the charge has to be considered a small
perturbation, $\veps\rightarrow 0$, and we want to use linear perturbation theory on
a FRW universe. The relevant equation at this purpose is
(\ref{maxwell1infconduct}): a constraint on the charge density can be derived
from constraints on the magnetic field and the vorticity induced by
it. We need to evaluate the time evolution
of these two quantities. At the lowest order, the
magnetic field evolution is
\be
\dot{B}_\al+\frac{2}{3}\Theta B_\al=0 \,,
\ee
and the vorticity evolution is \cite{TM,BDE} 
\be
\dot{\om}_{\al}+\frac{2}{3}\Theta\,\om_\al=
-\frac{1}{2}\ep_{\al\b\ga}D^\b\dot{u}^\ga\,.
\label{omegaevolcovariant}
\ee
Substituting with the momentum conservation equation
(\ref{monconservinfconduct}), and using the identity $\ep_{\al\b\ga}D^\b D^\ga f=-2\dot
f \om_\al$ \cite{TM}, equation (\ref{omegaevolcovariant}) becomes   
\be
\dot{\om}_{\al}+\left(\frac{2}{3}-w\right)\Theta\,\om_\al=
-\frac{1}{2\rho(1+w)}\ep_{\al\b\ga}D^\b {\ep^\ga}_{\de\mu}J^\de B^\mu\,.
\label{omegaevolJB}
\ee

The magnetic field evolution
equation gives the usual scaling behaviour $B_\al=B_{0\al}/ 
a^2$, where $B_0$ denotes the field amplitude today. 
The evolution for the vorticity is more involved. 
We know however that the charge density evolves like the
number density of particles, $\dot q+\Theta q=0$: this equation can
give us insight on the behaviour of vorticity. By deriving
Eq.~(\ref{maxwell1infconduct}), and imposing the charge scaling,
$2\dot \om_\al B^\al+2 \om_\al \dot B^\al=-\Theta q$, we find, from the
magnetic field evolution  
\be
B^\al \dot \om_\al=-\frac{1}{3}\Theta B^\al\om_\al \,.
\ee
This identity is satisfied in the two cases $(a\om_\al)^{\cdot}=0$,
and $(a\mbox{\boldmath$\omega$})^{\cdot}\perp {\bf B}$. In the first case, the evolution
of the vorticity is such that $\dot \om_\al+\frac{\dot
  a}{a}\,\om_\al=0$; in the second case, one has that only the component
parallel to the magnetic field satisfies $\dot \om_\parallel+\frac{\dot
  a}{a}\,\om_\parallel=0$. In the following, we will always assume that
$\om_\al\propto a^{-1}$. Given that we want to constrain the charge
using Eq.~(\ref{maxwell1infconduct}), we are in fact interested only
in $\om_\parallel$. However, in the following we will set constraints on the
vorticity vector independently, through the induced CMB anisotropies: 
our assumption, therefore, corresponds to postulating that the
entire vorticity contribution to CMB anisotropies comes only from the
$\om_\parallel$ component. A premise that is
conservative, in the sense that it makes the final limit on the charge 
less tight. 

We now proceed to evaluate the bounds on the charge in two different
configurations: uniformly and stochastically distributed charge. The 
strategy of limiting the charge asymmetry of the universe by using the observed
isotropy of the FRW spacetime has already been adopted in two previous
works: Orito and Yoshimura \cite{OY} also used the measurement of CMB
temperature fluctuations, instead Masso and Rota
\cite{MR} derived a bound using
Nucleosynthesis. In both these references, however, it 
is claimed that the presence of a non-zero
charge density would generate a large scale electric field in the
universe: they modelled the universe as an insulating medium rather
that a highly conducting one. Our work differs from both of them in
this, we believe, crucial aspect.

\section{Uniform distribution of charge}

We first first focus on a uniform charge distribution in a homogeneous 
spacetime. The anisotropic expansion of the universe will leave an imprint on
the propagation of light rays from the last-scattering surface
until today. 
In \cite{bfs} tight constraints
were derived on both $\omega$ and $B$ for a general class of
homogeneous, anisotropic models. It was found that
\begin{eqnarray}
B(\tau_0)&<& 3\times 10^{-9 }\Omega^{1/2}~h~{\rm Gauss}\,, \\
\omega(\tau_0)& < & 10^{-7}~H_0\,,\label{omegaconstrhomog}
\end{eqnarray}
where $H_0=100\,h \,{\rm km}\,{\rm sec}^{-1}{\rm Mpc}^{-1}$
is the Hubble constant today, and $1-\Omega$ is the curvature.
The bound in Eq.~(\ref{omegaconstrhomog}) was found assuming the usual
scaling for the vorticity in the matter era in the absence of sources: 
$\om_\al=\om_{0\al}/a^2$, as can be derived from
Eq.~(\ref{omegaevolJB}). Therefore, we have to correct this bound
accounting for our assumption for the evolution of vorticity sourced by the
charge: $\om_\al=\om_{0\al}/a$. This simply changes the limit
by a factor of $1/a(\eta_{\rm rec})\simeq 10^3$: $\omega_0< 10^{-4}~H_0$. 
With these constraints we obtain a limit on the charge asymmetry
using the Schwarz inequality on Eq. (\ref{maxwell1infconduct}), 
\bee q\leq
2|\omega(\tau_0)||B(\tau_0)|\leq 2.4\times
10^{-74}~h^2~\Omega^{1/2}\, {\rm GeV}^3\,. \label{q}\eee

It may be that the charge asymmetries corresponding to individual 
particle species are such as to cancel to some extent, yielding
to a universe which is more neutral overall. However, should this
not be the case, this last equation imply a maximal allowed charge
$q_X=q/n_X$ for each different constituent $X$, where $n_X$ is the
number density of charged particles.
For a flat universe ($\Omega=1$), and assuming that the charge
is due to a difference between the charges of the electron and the proton, we
divide by the baryon density of the universe $\Omega_B h^2=0.02$ to find 
\bee q_{e-p}\llsim 10^{-26} e \,. \label{qe-p}\eee

If the charged particles are instead dominated by
the dark matter constituents, we get $q_{\rm dm}\llsim 4\times 10^{-27} e 
\,\, m_{\rm dm}$ GeV$^{-1}$. For example, a conservative limit comes from 
neutralinos of mass of $1$ TeV, $q_{\rm dm}\llsim 4 \times
10^{-24} e$; considering instead a light dark matter particle with
$m_{\rm dm} \sim 10$ MeV \cite{celine1}, one gets $q_{\rm dm}\llsim 4 \times
10^{-30} e$. If we assume that the charge imbalance is predominantly caused by
photons, with $n_\gamma=421.84\,\, {\rm cm}^{-3}$, we
obtain 
$q_\gamma\llsim
10^{-35} e$; for each species of neutrinos we obtain instead $q_\nu\llsim
0.4 \times 10^{-34} e$, from $n_\nu=115.05\,\, {\rm cm}^{-3}$.
All these constraints are orders of magnitude more restrictive than 
the limits from laboratory experiments or astrophysical observations 
mentioned above, subject to the 
assumption that any charge asymmetries between species are not
anti-correlated.

\section{Stochastic distribution of charge}

One might expect that any mechanism of charge creation
will be local, possibly a result of local violation of
gauge invariance arising at some phase transition, or due to the
escape of charge into extra dimensions. It would then be more
appropriate to consider a stochastic distribution of charge
asymmetry with, or without, a net imbalance in charge.
The simplest assumption we can make, is that the magnetic field
and the vorticity created by the charge density are two independent
stochastic variables, with Gaussian distribution.
Homogeneity and isotropy forces the first moment of the distributions to
be zero and hence $\langle q \rangle = 0$. 
We further assume that the magnetic field and vorticity induced have power
law power spectra, $B^2(k)=B\, k^n$ and $\omega^2(k)=\Omega\, k^m$, where
the two spectral indices are different consequently to the 
independence assumption. We have then that: 
\bea
\langle B_i({\bf k}) B_{j}^*({\bf q}) \rangle &=& \frac{(2\pi)^3}{2}\delta^3({\bf k}-{\bf
q})(\delta_{ij}-\hat{k}_i\hat{k}_j)\,B^2(k)\,,\:\:\:k<k_c \nonumber \\
\langle \omega_i({\bf k}) \omega_{j}^*({\bf q}) \rangle &=&
\frac{(2\pi)^3}{2}\delta^3({\bf k}-{\bf 
q})(\delta_{ij}-\hat{k}_i\hat{k}_j)\,\omega^2(k)\nonumber\,,
\eea
   
where we need to introduce an upper cutoff frequency $k_c$ in the magnetic
field power spectrum, which accounts for the interaction of the magnetic
field with the cosmic plasma at small scales; $k_c(\tau_{\rm rec})\simeq
10\,{\rm Mpc}^{-1}$, 
from \cite{SB}. For the following calculations, we do not need to
assume a cutoff frequency for the vorticity. The factors
$(\delta_{ij}-\hat{k}_i\hat{k}_j)$ come from the divergence-free
property of both the magnetic field and the vorticity \cite{BDE}. 
We define $B_{\lambda}^2 =\langle B^i({\bf x})B_i({\bf x})\rangle
|_{\lambda}$ and $\omega_{\lambda}^2 =\langle
\omega^i({\bf x})\omega_i({\bf x})\rangle |_{\lambda}$ 
to be the energy densities of the magnetic field and the vorticity
in a region of size $\lambda$. 

We wish to estimate the mean fluctuation of charge in a region of
size $\lambda$:
\bea \langle
\,q^2_{\lambda}\,\rangle=\frac{1}{V_\lambda^2}\int d^3r_1 d^3r_2\,
e^{-\frac{r_1^2} {2\lambda^2}}\, e^{-\frac{r_2^2}{2\lambda^2}}
\langle q({\bf r}_1+{\bf x})q({\bf r}_2+{\bf x}) \rangle\,.
\nonumber \eea 
From Eq.~(\ref{maxwell1infconduct}) one has 
\be q({\bf k})=\frac{2}{(2\pi)^3}\int d^3 p\,
\omega_i({\bf k}-{\bf p})B^i({\bf p})\,, \ee
so the charge density power
spectrum is
\be
\langle q({\bf k}) q^*({\bf h})\rangle =  
\Omega B \,\delta^3({\bf k}-{\bf h})
\int d^3 p \,|{\bf k}-{\bf p}|^m\,p^n
[1+(\widehat{\bf k-p}\cdot\hat{\bf p})^2]\nonumber\,,
\ee 
with $0<p<k_c$. We obtain an analytical approximation:
\bee
\langle\,q^2_{\lambda}\,\rangle \simeq
\frac{4\,B_{\lambda}^2\,\omega_{\lambda}^2}
{\Gamma(\frac{n+3}{2})\,\Gamma(\frac{m+3}{2})}\,f(n,m,\lambda,k_c)\label{qlim}\,,
\eee
with 
\bea
f&=&\frac{2\,(\lambda k_c)^{n+3}}{3\,(n+3)}\,\Gamma\left(\frac{m+3}{2},(\lambda
k_c)^2\right)+ \frac{(\lambda k_c)^{n+m+3}}{n+m+3}\times\nonumber\\
&&\left[\Gamma\left(\frac{3}{2}\right)
-\Gamma\left(\frac{3}{2},(\lambda k_c)^2 \right)\right] 
+\frac{2m-n-3}{3(n+3)(n+m+3)}\times\nonumber\\
&&\left[\Gamma\left(\frac{n+m+6}{2}\right)
-\Gamma\left(\frac{n+m+6}{2},(\lambda k_c)^2 \right)\right]\,, \nonumber 
\eea
where $\Gamma(a,x)$ denotes the incomplete gamma function (6.5.3 of \cite{AS}).

Eq.~(\ref{qlim}) provides a limit for the charge asymmetry
parameter $q_X=\sqrt{\langle \,q_\lambda^2 \,\rangle}/ n_X$ today, as a function of the
parameters characterising the magnetic field and the vorticity:
amplitudes, spectral indices, cutoff frequency and coherence scale.

The amplitude $B_\lambda$ of a cosmic magnetic field must satisfy stringent
constraints, which have been found in previous works using
completely independent methods. The most direct constraint on 
the strength of a magnetic field comes from
the observation of Faraday rotation in radio sources:
for example, in \cite{RB} it is shown that, for a cluster magnetic field,
$B_\lambda\llsim 10^{-6} {\rm Gauss}$ on scales $\lambda\simeq 1\,{\rm kpc}$.
If the charge imbalance, and consequently the magnetic
field, was created before recombination, we can apply the limits
which come from CMB observations (see for example \cite{JKO}): 
in \cite{DFK} limits where derived, by considering
the anisotropies induced in the CMB by gravitational waves, created
by the anisotropic stresses of a primordial magnetic field (see Eq.~(28) of \cite{DFK}, for
$\lambda=0.1 h^{-1} {\rm Mpc}$). Finally, in \cite{CD}, a constraint was
found on the amplitude of a magnetic field created before
Nucleosynthesis, as a function of the spectral index $n$, by imposing
the Nucleosynthesis bound on the energy
density of the gravitational waves induced again by the field (Eq.~(33) of
\cite{CD}, also for $\lambda=0.1 h^{-1} {\rm Mpc}$).

We next need to find a constraint on $\omega_\lambda$, which is not present in the
literature. Again, it can be found using the isotropy of the CMB. We
need to evaluate the vector contribution to the CMB temperature
anisotropy, and we use the formalism of \cite{HW}. 
Our vorticity variable $\om_i=\frac{1}{2} \ep_{ijk}\,\om^{jk}$  
is related to the vector gauge invariant variable
$V_C=v\har{1}-B\har{1}$, where $v\har{1}$ is the vector part of the
fluid velocity perturbation and
$B\har{1}$ the vector metric perturbation,
through relation 
$\om_{ij}=a\,V_C\,Q^{(1)}_{[i|j]}$, where $Q^{(1)}_{i}$ denotes the
vector harmonic \cite{BDE}. Therefore, $\om_i=\frac{k}{2}(v\har{1}-B\har{1})Q\har{1}_i$.  
For the purposes of this analysis, we neglect the finite thickness of the
last scattering surface and work in the tight coupling limit, for which
$v\har{1}=v_\ga=v_B$. In this limit, the dominant effect of vector
perturbations is the creation of a dipole in the temperature
anisotropy. From the general expression for the CMB power spectrum from
vector perturbations given in \cite{HW} we finally find 
\bee
C_\ell=\frac{8}{\pi} \frac{\ell(\ell+1)}{a^2(\tau_{\mathrm{rec}})}
\int_0^\infty dk\,\frac{j_\ell^2(k\tau_*)}{(k\tau_*)^2}\,\omega^2(k)\,, 
\eee  
where $\tau_*=\tau_0-\tau_{\mathrm{rec}}$, and the factor 
$a^{-2}(\tau_{\mathrm{rec}})$ accounts for the vorticity time
evolution. This gives us 
\bee
\frac{\ell^2 C_\ell}{2\pi}\simeq 4\sqrt{\pi}
\frac{\Gamma(\frac{3-m}{2})}{\Gamma(\frac{m+3}{2})\Gamma(\frac{4-m}{2})} 
\,z_{\mathrm{rec}}^2\,\omega_\lambda^2\,
\frac{\lambda^{m+3}}{{\tau_*}^{m+1}}\,\ell^{m+1}\,. 
\eee
The maximum value for the CMB temperature
anisotropy is  $\ell^2 C_\ell/(2\pi)\simeq 10^{-10}$, and in order to
constrain $\omega_\lambda$ we fix the $\ell$
dependence by considering the two extreme values $\ell=4$, if $m<-1$,
and $\ell=200$, if $m>-1$.     

\begin{figure}
\centering
\resizebox{11cm}{12.5cm}{\includegraphics{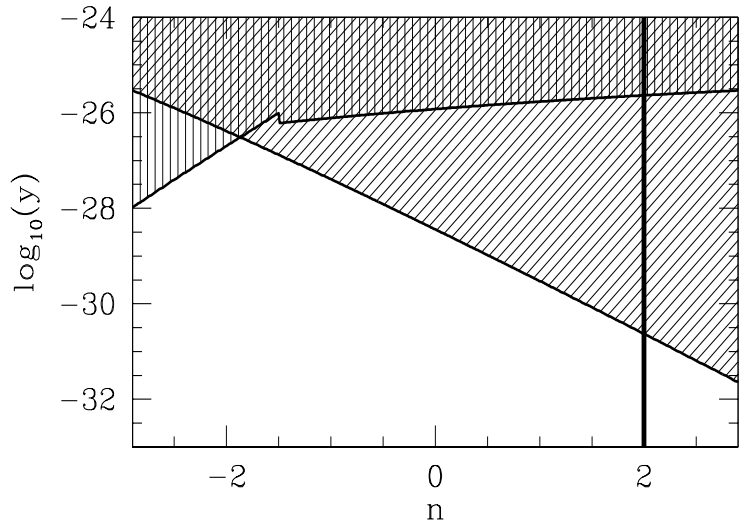}}
\vspace{-1.8in}
\caption{This figure shows the limits on the charge fraction 
  parameter $y$ which apply to a stochastic distribution of charge, as a function
  of the spectral index of the induced magnetic field, for $\lambda=0.1 h^{-1} {\rm Mpc}$. 
  The diagonally-shaded region represents
  the exclusion region derived from nucleosynthesis  
  constraints~\protect\cite{CD}; the 
  vertically-shaded region is excluded by CMB anisotropies~\protect\cite{DFK}. 
  Magnetic field constraints from
  clusters~\protect\cite{RB} give a less restrictive bound,
  which is not shown in the figure. If the process generating the
  charge is causal, then $n \geq 2$.
\label{fig1}}
\end{figure}

We can finally evaluate Eq. (\ref{qlim}), and derive an upper bound on
the charge density in a region of size $\lambda$, by applying the aforementioned
constraints on $B_\lambda$ and the limit on $\omega_\lambda$ just derived.
The bounds 
are represented in fig.~\ref{fig1}, as a function of the spectral index
$n$ of the magnetic field. For every value of $n$, we have maximised
(\ref{qlim}) with respect to the vorticity spectral index $m$, on
which we have no theoretical prediction.   
Note that the quantity shown in the y-axis, 
\bee
y=\frac{q_X}{e}\left(\frac{\Omega_X}{\Omega_B}\frac{m_P}{m_X}\right)
\eee
is the charge fraction parameter `normalised' to the case
of `baryonic' charge asymmetric matter (as in Eq.~(\ref{qe-p})). If the process generating the
charge and the magnetic field is causal, \emph{e.g.} it takes place
during a phase of standard Friedmann expansion, then the spectral
index must satisfy $n\geq 2$ \cite{DC}. In this case, the limits on
$q_X$ are more stringent than in the case in which the generation
occurs during inflation. 

\section{Discussion} 

In this paper we have derived a cosmological constraint on the
presence of a non-zero electric charge in the universe. 
A conservation law associated with a long range force, such as charge
conservation with the electromagnetic force, 
is usually believed to be exact as a result of gauge
invariance. Conservation of charge appears to hold in all particle
decays, and there are strong experimental constraints on 
the charge of particles which
are predicted to be neutral by the standard model. 
However, even though electric charge conservation is well established
on Earth, this may not imply directly the overall
neutrality of the universe: to be able to draw this conclusion, 
one would have to assume in addition that
charge has to be conserved on all scales and during the entire
evolution of the universe. This may not be the case and
 the determination of a cosmological constraint on the charge
asymmetry of the universe is of conceptual importance. 

There are a host of proposals
that lead to the generation of a charge asymmetry in the
universe. In the context of brane world scenarios, charge
non-conservation arises in the form of charge leakage
into extra dimensions; in varying speed of light theories, it can be 
a consequence of the variation of the fine structure constant;
allowing for some 
modification of the standard model, it can arise if 
the gauge symmetry is temporarily broken during a phase transition
taking place in the early universe, or if one imposes a small,
non-zero mass to the photon. In our analysis, 
we did not consider a complete model in which the charge
asymmetry originates: by doing so, our constraints gain in generality. 
The assumption we made, that charge is conserved in the period in which our
constraint is established, points toward a model in which the charge
is created during a primordial phase transition leading to 
a transitory breaking of the 
electromagnetic gauge symmetry. 

We have obtained our constraints on the overall charge imbalance of
the universe  
in two different cases: a uniform
distribution of charge, and a distribution characterised by stochastic
fluctuations. In order to derive our results, we made the assumptions  
that the universe is a good conductor, and that the charge is a first
order perturbation in a background FRW model; moreover, we
generalised the scaling of the charge-induced vorticity as the inverse of the
scale factor. These are the only assumptions which affect the
uniform distribution limit; in the case of the stochastic
distribution, we further assumed that vorticity and magnetic field are
two independent, Gaussian stochastic variables. A possible 
extension of our analysis would be to go beyond the linear calculation,
and evaluate the full effect that the presence of a non-zero charge 
has on the background dynamics of the spacetime. 

In the case of a uniform distribution of charge, it is remarkable that
our cosmological limit, once translated in terms of constraints for
the single charge carriers, gives results which are orders of
magnitude stronger than the ones derived by terrestrial experiments or
astrophysical observations. If the charge is distributed
stochastically, the limits are less stringent: this is a consequence
of the fact that the CMB bound on stochastic vorticity is
less tight than in the uniform case, and in addition we have maximised it with
respect to the vorticity spectral index $m$. However, we still find interesting
constraints for high values of the magnetic field spectral index:
these would apply, in the case of a causally created charge asymmetry,
for example during a phase transition.

\ack
We are very grateful to Steven Biller for suggesting this project and
for fruitful discussions. 
We thank Christos Tsagas, John Barrow, Ruth Durrer, Roy Maartens 
and Katherine Blundell for useful conversations.
PGF acknowledges support from the Royal Society. CC research is
funded through the generosity of the Dan David Prize Scholarship 2003.

\end{document}